# On multi-scale percolation behaviour of the effective conductivity for the lattice model with interacting particles

R. Wiśniowski[a,1], W. Olchawa[b], D. Frączek[c,1], R. Piasecki[d,*],

[a,b,d] *Institute of Physics, University of Opole, Oleska 48, 45-052 Opole, Poland*
[c] *Department of Materials Physics, Opole University of Technology, Katowicka 48, 45-061 Opole, Poland*

## HIGHLIGHTS

- We study the multi-scale site percolation in a system with interacting objects.
- The interaction energy modifies the value of the percolation threshold in a limited range.
- When a repulsive interaction dominates thermal energy, the exact $\phi_c$ at scales $k = 2, 3$ is found.
- In turn, when $k \gg 1$, the evaluated formula for $\phi_c(k)$ implies the limit threshold value 0.75.

## ABSTRACT

Recently, the effective medium approach using $2 \times 2$ basic cluster of model lattice sites to predict the conductivity of interacting droplets has been presented by Hattori *et al*. To make a step aside from pure applications, we have studied earlier a multi-scale percolation, employing any $k \times k$ basic cluster for non-interacting particles. Here, with interactions included, we examine in what way they alter the percolation threshold for any cluster case. We found that at a fixed length scale $k$ the interaction reduces the range of shifts of the percolation threshold. To determine the critical concentrations, the simplified model is used. It diminishes the number of local conductivities into two main ones. In the presence of a dominance of the repulsive interaction over the thermal energy, the exact percolation thresholds at scales $k = 2$ and $3$ can be obtained from analytical formulas. Furthermore, by a simple reasoning, we obtain the limiting threshold formula for odd $k$. When $k \gg 1$, the odd-even difference becomes negligible. Hence, the 0.75 is the highest possible value of the threshold.

(Some figures in this article are in colour only in the electronic version)

[*] Corresponding author. Tel.: +48 77 452 7285; fax: +48 77 452 7290.
E-mail addresses: ryszard.wisniowski@gmail.com (R. Wiśniowski), wolch@uni.opole.pl (W. Olchawa), d.fraczek@po.opole.pl (D. Frączek), piaser@uni.opole.pl (R. Piasecki).
[1] R. Wiśniowski and D. Frączek are recipients of a Ph.D. scholarship under a project funded by the European Social Fund.

## 1. Introduction

One of the simple techniques to investigate essential physical properties of random heterogeneous materials is the network extension of effective medium approximation (EMA). A few years ago, Hattori *et al*. presented a lattice EMA-model that describes electrical percolation in microemulsion solution with interacting droplets [1]. Instead of the $2 \times 2$ basic cluster of lattice sites used in Ref. [1], the earlier modification of this model takes into account any cluster's size $k \times k$, yet for zero interaction energy [2]. It was found that an increase in the basic cluster's size shifts the critical concentration close to one. Such an approach can be treated as a high-temperature description of the initial lattice model, see Appendix in [2].

The inclusion of interactions between two adjacent conducting particles and accounting for the related chemical potential is the next step in the proposed multi-scale approach to the lattice model. Using the same notation as in Ref. [1], we consider the physically sound range of interaction energies. On the one hand, the attractive interaction with $0 < \beta \Delta \leq 1$ is taken into consideration, where the standard notation $\beta \equiv 1/k_B T$ is applied. On the other hand, the more important repulsive case with $-\infty < \beta \Delta < 0$ is also examined.

It is worth remarking that the interacting particles are treated as non-overlapping finite-size objects, that is as unit square cells centred on the lattice sites [1]. Frequently, in computer simulations, pixels represent the smallest objects. This allows considering effective properties/microstructure relations, making use of digitized images of real samples [3-5]. However, we should bear in mind that the effective conductivity and percolation itself could be influenced by the size of mono-grains [6] as well as by the grain size distribution in a random material [7] or polydispersity [8]. The usage of the finite size objects can also be extended to percolation of dimers on square lattices [9] and further, to the impact of defects on percolation in random sequential adsorption of linear $k$-mers on square lattices [10]. Along this line, but for a three-dimensional (3D) system of straight rigid rods of length $k$, the ratio between percolation threshold and jamming coverage ($\theta_p/\theta_j$) shows a non-universal behaviour, monotonically decreasing to zero, with increasing $k$ [11]. In turn, for a system consisting of overlapping squares or cubes, the percolation behaviour of obstacles and of the void space was carefully analysed as well as the transition between continuous and discrete percolation [12]. This shortened list of papers making use of finite-size objects in the context of percolation ends with the work devoted to 3D electric conductivity of overlapped spheroids of revolution in continuum [13]. The conductivity curves depend upon the shape of conductive spheroids while the percolation threshold shows a non-monotonic dependence on the related aspect ratio. It should be mentioned that interactions cause additional correlations between finite-size objects. This is another interesting task in statistical physics.

Here, we concentrate on two queries: *i*) in what way at *fixed* scales the strength of the interactions modifies the system's percolation threshold and reversely, *ii*) how a given interaction energy of adjacent particles influences the percolation behaviour at *different* length scales *k*. In that way, the work completes the earlier paper [2].

## 2. The model Z4 and its simplified version Z4s

Let us consider a scale extension of the two-dimensional regular lattice EMA-based model with *M* sites used by Hattori *et al*. [1]. For a unit lattice-distance, square cells 1×1 are centred on the lattice sites. Each of them can be occupied by a particle ($s_i = 1$), here represented by a black pixel of the higher conductivity *H*-phase with the global volume fraction $\phi$ and a given electrical conductivity $\sigma_H \gg \sigma_L$. For the *L*-phase with the global volume fraction $\theta \equiv 1 - \phi$, its $\sigma_L$-conductivity is attributed to an empty cell ($s_i = 0$) represented here by a white pixel. Following the paper [1], we suppose that Hamiltonian *H* of the system has the form:

$$\mathcal{H} = -\Delta \sum_{<ij>} s_i s_j - \mu \sum_{i=1}^{M} s_i \ . \tag{2.1}$$

Here $\Delta$ is the interaction energy between two adjacent black pixels, $\mu$ is the chemical potential of *H*-particle and summation on $<ij>$ is over the nearest neighbour sites.

Inspired by the assumption used in [1], about neglecting fluctuations of electrical potential in the orthogonal direction to the macroscopic electric field directed along *x*-axis, we focus on investigation of a possible multi-scale behaviour of electrical percolation in such a model. Thus, modifying further the Z4-model [2], we consider a general case of an elementary $k \times k$ cluster in the presence of interactions. Now, the Z4-model accounts for all cluster conductivities and energies.

At this stage, our main task is to obtain, at any fixed length scale *k*, the probability of appearance of a class of configurations $C_i(k)$, each of the local conductivity $\sigma_{x,i}(k) \equiv \sigma_i$ computed according to the formula

$$\sigma_i = \sum_{j=0}^{k} \frac{r_j}{j/\sigma_H + (k-j)/\sigma_L} \ . \tag{2.2}$$

Here, *k* denotes the number of rows, while the quantity $r_j \in \{r_0, \ldots, r_j, \ldots, r_k\}$ describes the number ofrows with exactly *j* black pixels representing *H*-phase. It should be noticed that the local conductivity given by Eq. (2.2) is independent of the permutations of pixels in any row and of the permutations of rows, either, while the cluster energies are dependent on them.

We remind that the following obvious conditions should also be satisfied:

$$r_0 + r_1 + \cdots + r_k = k \tag{2.3a}$$

and

$$r_1 + 2r_2 + \cdots + k r_k = n_i, \tag{2.3b}$$

where $0 \leq n_i \leq k^2$ is a number of black pixels of the considered cluster. Now, we are in a position to give the needed formula for the probability $p_i(k) \equiv p_i$ of the appearance of a class of configurations having a local conductivity $\sigma_i$ and with attributed parameters $m_i$, $n_i$

$$p_i = g_i \frac{e^{\beta(m_i \Delta + n_i \mu)}}{\mathscr{Z}}, \tag{2.4a}$$

where the partition function $\mathscr{Z} \equiv \mathscr{Z}(k)$ is determined from the condition

$$\sum_i p_i = 1, \tag{2.4b}$$

while the chemical potential $\mu$ must fulfil the following condition

$$\sum_i n_i p_i = k^2 \phi. \tag{2.4c}$$

Here, $g_i$ stands for geometrical degeneracy multiplier and $m_i$ describes the number of pairs of interacting $H$-particles. From now on, each of $i$th class of configurations will be characterised by a triplet of the parameters: $\{\sigma_i, m_i, n_i\}$.

At any fixed scale $k$, all the possible classes of configurations, i.e. those with identical parameters $\sigma_i$, $m_i$ and $n_i$, are collected in Tab. 1. Column 2 shows the numbers describing all configurations of a system; Column 3 – the configurations grouped into classes; then Column 4 includes their splitting into $H$-configurations with $r_k > 0$ and Column 5 – the rest $L$-one with $r_k = 0$. The last two columns become important when we introduce a simplification of the Z4-model.

Tab. 1.

| $k$ | # of config. = $2^{k^2}$ | # of classes | # of $H$-classes | # of $L$-classes |
|---|---|---|---|---|
| 2 | 16 | 7 | 3 | 4 |
| 3 | 512 | 55 | 24 | 31 |
| 4 | 65536 | 386 | 174 | 212 |
| 5 | 33554432 | 2503 | 1168 | 1335 |

Following the idea of the effective medium, a hypothetical regular network is considered, where the conductance of the resistor on each bond is of the same value. Within the well-known symmetrical effective medium theory of Bruggeman, see, e.g. [14], the 2D effective conductivity $\sigma^*$ at a fixed length scale is given by

$$\sum_i p_i \frac{\sigma_i - \sigma^*}{\sigma_i + \sigma^*} = 0. \tag{2.5}$$

To present the numerical results for a two-phase system, the higher conductivity component, $\sigma_H = 10^6$, and a lower one, $\sigma_L = 1$ in a.u. are assumed. In Fig. 1a, the $k$-multi-scale dependence of the total effective conductivity $\sigma^*(k; Z4)$ as a function of volume fraction $\phi$ is demonstrated for chosen scales, $k = 2, 3$ and 5, and two fixed interaction energies: $\beta\Delta = 1$ (attractive case) and $\beta\Delta = -1$ (repulsive one). The displacements of the percolation threshold $\phi_c(k)$ along the increasing $k$-scales can be observed for both interactions. In turn, the open circles indicate the case of non-interacting particles when $\beta\Delta = 0$.

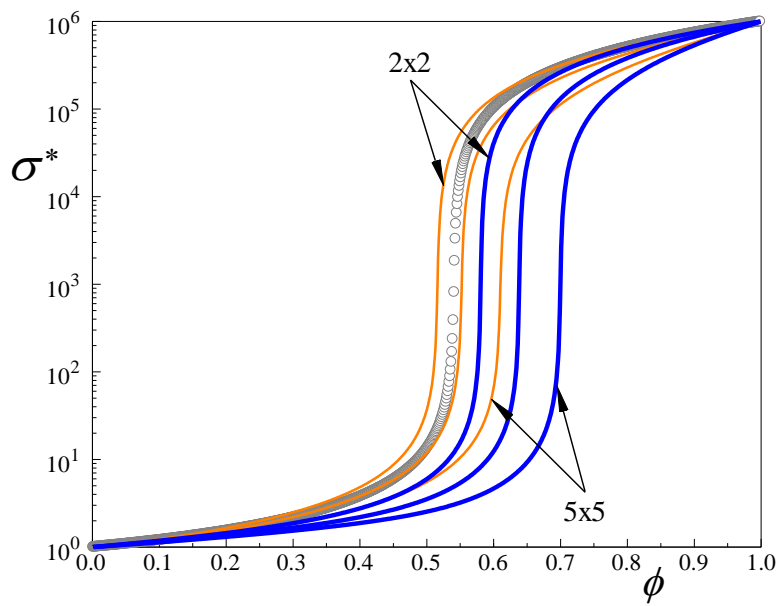

**Fig. 1a.** The multi-scale percolation behaviour of the effective conductivity $\sigma^*(k; Z4)$ as a function of the volume fraction $\phi$ of highly conducting $H$-phase for the chosen attractive, $\beta\Delta = 1$ (the thin orange lines) and the repulsive, $\beta\Delta = -1$ (the thick blue lines) interaction at exemplary length scales, $k = 2, 3$ and 5. The conductivity of $H$-phase and $L$-phase is $10^6$ and 1 in a.u., respectively. The shifting of the percolation threshold is clearly seen along the increasing $k$-scales for both interactions. Notice that for a fixed length scale, e.g. $k = 2$ (the upper arrows), the shifts in opposite directions appear for attractive and repulsive interactions, on the left and on the right compared to the case of non-interacting particles, when $\beta\Delta = 0$ (the open circles).

The reverse situation relating to a fixed scale $k = 3$ and for various values of $\beta\Delta = 1$ (the thin orange line), 0 (the open circles), $-1, -4$ and $-10$ (the thick blue lines) is illustrated in Fig. 1b. Now, the shifting of the percolation threshold appears within the limited range of the volume fraction $0.5 \leq \phi_c < \phi_c^* = 13/18$; the justification of this threshold value will be given further. The rightmost thick blue line ($\beta\Delta = -10$) practically approaches the numerical limit when $\beta\Delta \to -\infty$ for the repulsive interaction. Such a behaviour may result from a geometrical reasons considered earlier, when for non-interacting particles a change in the basic cluster size shifts the $\phi_c$ up to one [2].

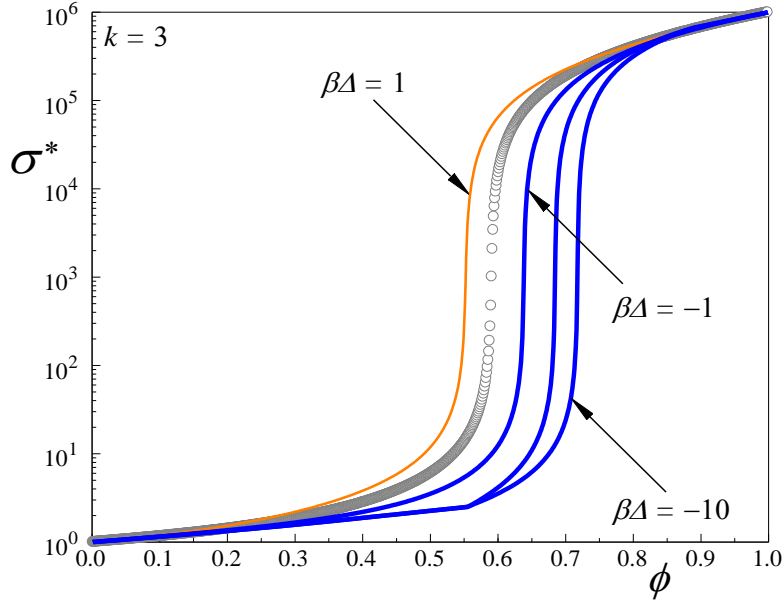

**Fig. 1b.** Similarly as in Fig. 1a, but for the fixed length scale $k=3$ and various values of $\beta\Delta=1$ (the thin orange line), 0 (the open circles), $-1$, $-4$ and $-10$ (the thick blue lines). Now, for repulsive interactions the shifting of the percolation threshold can be observed within the limited range of the volume fraction $\phi$, i.e. the rightmost thick blue line practically approaches the numerical limit, when $\beta\Delta \to -\infty$.

We would like to remark that an accurate evaluation of critical concentrations at different scales and interaction energies can be obtained easily by means of the simplified version of Z4-model, referred to as Z4s-model here. Such an evaluation is available due to an essential reduction of the whole number of classes into two basic ones, independently of the length scale (see the last two columns in Tab. 1). The Z4s-model employs only the $\sigma_H$ and $\sigma_L$ conductivities attributed to the two separate sub-classes: the so-called *H*-class (with $r_k > 0$) and the *L*-class (with $r_k = 0$). The following approximations can be used:

$$\sigma_i \approx \begin{cases} r_k\,\sigma_H \approx \sigma_H & \text{for } r_k > 0 \\ \sigma_L \sum_{j=0}^{k-1} \dfrac{r_j}{(k-j)} \approx \sigma_L & \text{for } r_k = 0 \end{cases} \quad (2.6)$$

Within this simplified approach, the probability $p_H$ of the appearance of a configuration belonging to *H*-class for a given fraction $\phi$ can be written as

$$p_H(\phi;\,\beta\Delta,\,k) = \sum_{i,\,r_k > 0} p_i\,. \quad (2.7)$$

For the complementary *L*-class, the related probability $p_L$ simply equals

$$p_L(\phi;\ \beta\Delta,\ k) = 1 - p_H(\phi;\ \beta\Delta,\ k) \tag{2.8}$$

Consequently, the general formula (2.5) for the effective conductivity $\sigma^*$ simplifies to

$$p_H \frac{\sigma_H - \sigma^*}{\sigma_H + \sigma^*} + p_L \frac{\sigma_L - \sigma^*}{\sigma_L + \sigma^*} = 0. \tag{2.9}$$

Now, the natural question arises: How do the increase in the strength of interactions and the changes in the length scale alter the probabilities $p_H(\phi;\ \beta\Delta,\ k)$ and $p_L(\phi;\ \beta\Delta,\ k)$, which are described by Eqs. (2.7-8)?

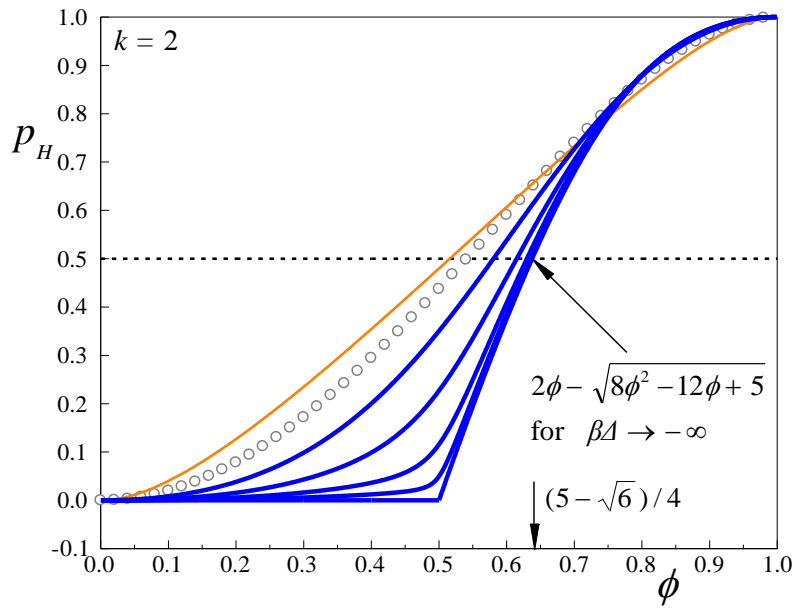

**Fig. 2a.** The probability $p_H(k; Z4s)$ of the appearance of *H*-class with the local conductivity of order $10^6$ at the fixed length scale $k = 2$ and for chosen values of $\beta\Delta = 1$ (the thin orange line), 0 (the open circles), $-1, -2, -3, -4$ and the limiting curve (thick blue) that is indicated by the upper arrow. For repulsive interactions, when $\beta\Delta \to -\infty$, the analytical limiting formula accessible for $p_H^*(k = 2; Z4s)$, see Eq. (2.15), is shown below the arrow. For better visibility, the $p_L(k; Z4s) = 1 - p_H(k; Z4s)$ corresponding to *L*-configurations with the local conductivity of order 1 are not shown. For the Z4s-model, the $\phi$-coordinate of the so-called balance point, when $p_H(k; Z4s) = p_L(k; Z4s) = \frac{1}{2}$ corresponds to the critical volume fraction $\phi_c(k)$ described in the text. The vertical arrow points on the limiting critical concentration $\phi_c^*(k = 2)$.

Similarly as in [2], the $p_H$ and $p_L$-curves intersect at the so-called balance point, which indicates approximately the pertinent critical concentration $\phi_c$ or equivalently, the relevant percolation threshold. Thus, at any scale, the position of a percolation threshold $\phi_c$ can be determined from the equality

$$p_H(\phi_c;\ \beta\Delta,\ k) = p_L(\phi_c;\ \beta\Delta,\ k) \equiv 1/2. \tag{2.10}$$

However, because of the quickly increasing numbers of configurations (see the second column of Tab. 1), we restrict ourselves to the length scales $k \leq 5$. Obviously, it is sufficient to consider intersections of the $p_H(\phi;\ \beta\Delta,\ k)$-curves with the horizontal line corresponding to the value of 0.5. In Fig. 2a, we present the probability $p_H(\phi;\ \beta\Delta,\ k)$ at the fixed length scale $k=2$ and for chosen values of $\beta\Delta = 1$ (the thin orange line), 0 (the open circles), $-1$, $-2$, $-3$, $-4$ and the limiting curve (the thick blue one) that is distinguished by the arrow. On the other hand, the related limiting critical concentration and details of its obtaining will be given below.

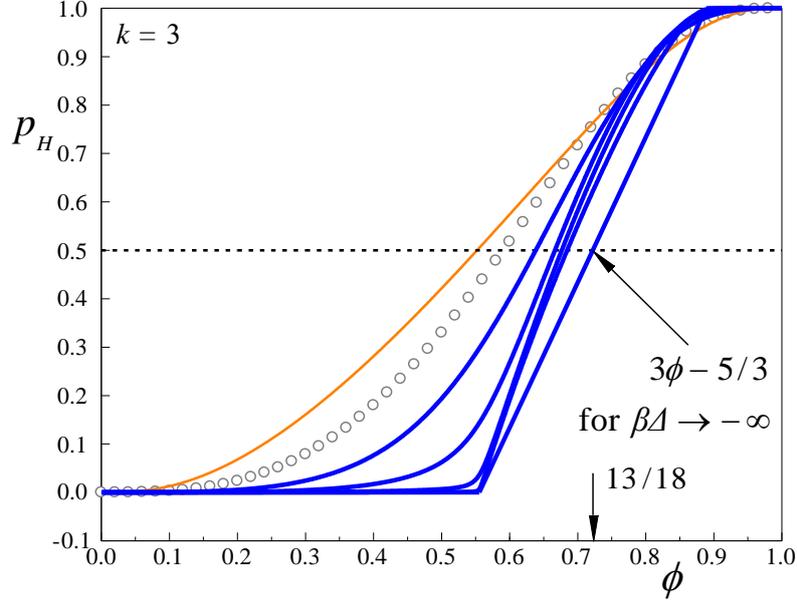

**Fig. 2b.** The same as in Fig. 2a but at the fixed length scale $k=3$ for chosen values of $\beta\Delta = 1$ (the thin orange line), 0 (the open circles), $-1$, $-2$, $-3$, $-4$ and the limiting curve (thick blue) indicated by the upper arrow. Again, for repulsive interactions the limiting exact formula is available this time for $p_H^*(k=3; \text{Z4s})$, see Eq. (2.16). Similarly as in Fig. 2a, the curve corresponding to $\beta\Delta = -10$ is hardly distinguishable from the limiting one.

The similar situation as in Fig. 2a, but at the fixed length scale $k=3$ with the same chosen values of $\beta\Delta = 1$ (the thin orange line), 0 (the open circles), $-1$, $-2$, $-3$, $-4$ and the limiting curve (the thick blue one), is presented in Fig. 2b. Again, the analytical formula is available for the limiting probability $p_H$ when $\beta\Delta \to -\infty$ for the repulsive interaction. Likewise in Fig. 2a, the curve corresponding to $\beta\Delta = -10$ is hardly distinguishable from the limiting one.

The above two figures suggest that one can expect an appearance of limiting critical concentrations for an extremely strong repulsive interaction. Let us try to obtain the following limit for any fixed scale $k$

$$p_H^*(k) \equiv p_H(\phi;\, \beta\Delta \to -\infty,\, k) = \lim_{\beta\Delta \to -\infty} \sum_{i,\, r_k > 0} p_i \, . \qquad (2.11)$$

Indeed, at a few initial scales one can obtain the analytical formula for $k=2, 3$, or its numerically estimated values for $k=4$ and 5. However, when $k \geq 6$, it turns out that under simple assumptions a reliable evaluation of the critical concentration can be found; see the formula given below by Eq. (2.18). The corresponding results are described by the Eqs. (2.15-18).

We remind that the limiting value of $\phi_c(k)$, i.e. when $\beta\Delta \to -\infty$, is denoted as $\phi_c^*(k)$. Specifically, in the simplest case $k=2$, we obtain

$$p_H^*(k=2) = \lim_{\beta\Delta \to -\infty} \frac{1}{\mathscr{Z}} \left( \phi^4 e^{4\beta(\Delta+\mu)} + 4\phi^3\theta e^{\beta(2\Delta+3\mu)} + 2\phi^2\theta^2 e^{\beta(\Delta+2\mu)} \right) \qquad (2.12)$$

and

$$\lim_{\beta\Delta \to -\infty} \frac{1}{\mathscr{Z}} \left( 4\phi^4 e^{4\beta(\Delta+\mu)} + 12\phi^3\theta e^{\beta(2\Delta+3\mu)} + 8\phi^2\theta^2 e^{\beta(\Delta+2\mu)} + 4\phi^2\theta^2 e^{2\beta\mu} + 4\phi\theta^3 e^{\beta\mu} \right) = 4\phi$$

(2.13)

where the partition function $\mathscr{Z}(k=2)$ is given by

$$\mathscr{Z} = \phi^4 e^{4\beta(\Delta+\mu)} + 4\phi^3\theta e^{\beta(2\Delta+3\mu)} + 4\phi^2\theta^2 e^{\beta(\Delta+2\mu)} + 2\phi^2\theta^2 e^{2\beta\mu} + 4\phi\theta^3 e^{\beta\mu} + \theta^4 \qquad (2.14)$$

The corresponding final form of the limiting probability reads

$$p_H^*(k=2) = \begin{cases} 0, & 0 \le \phi < \frac{1}{2} \\ 2\phi - \sqrt{8\phi^2 - 12\phi + 5}, & \frac{1}{2} \le \phi \le 1 \end{cases} \qquad (2.15)$$

and using Eq. (2.10), the limiting critical concentration can be computed, $\phi_c^*(k=2) = (5-\sqrt{6})/4$; see the vertical arrow in Fig. 2a. In the same figure, the corresponding limiting curve (the thick blue one) is distinguished by the upper arrow.

Unexpectedly, in the second case, i.e. the length scale $k=3$, the computations are much simpler. Note that now only two configurations among those (55) represented in Tab. 1 have non-zero probabilities when $\beta\Delta \to -\infty$. The first of them is a chessboard array with the black centre that belongs to the *L*-class (let us denote it as HL) while the second one is a wholly black configuration and it belongs to the *H*-class (let us mark it as HH). All that considerably simplifies the computations. Finally, the following analytical formula can be reached

$$p_H^*(k=3) = \begin{cases} 0, & 0 \le \phi < \frac{5}{9} \\ 3\phi - \frac{5}{3}, & \frac{5}{9} \le \phi \le \frac{8}{9} \\ 1, & \frac{8}{9} < \phi \le 1 \end{cases} \qquad (2.16)$$

Now, from Eq. (2.10) the limiting critical concentration results, $\phi^*_c(k=3) = 13/18$; see the vertical arrow in Fig. 2b. The corresponding limiting curve (the thick blue one) is indicated by the upper arrow in the same figure.

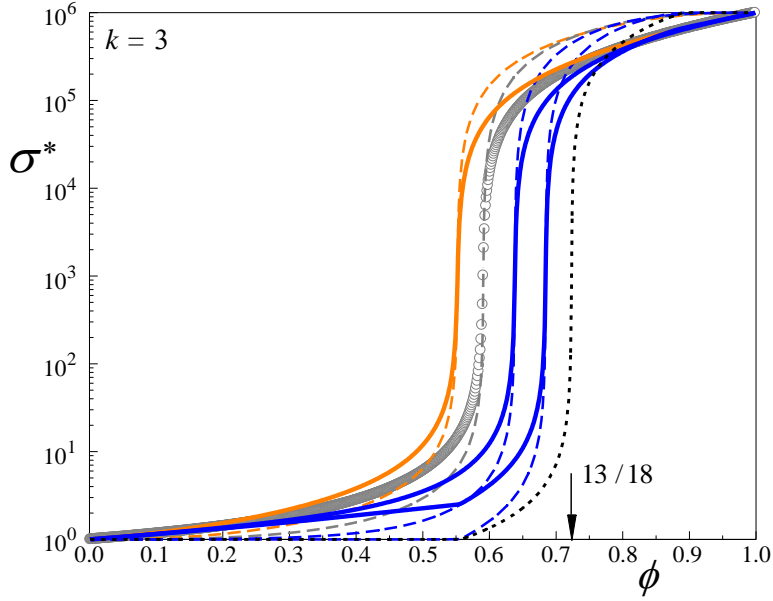

**Fig. 3a.** A comparison between the effective conductivity $\sigma^*(k; Z4)$, the thick solid lines, and $\sigma^*(k; Z4s)$, the thin dashed lines, at an exemplary length scale $k=3$ for the chosen values of $\beta\Delta = 1$ (orange), 0 (the grey open circles and the dashed line), $-1$, $-4$ (blue) and the limiting analytical curve $\sigma^*(k = 3; Z4s)$, the black short-dashed line when $\beta\Delta \to -\infty$. The arrow points on the highest (at this scale) accessible value of the critical volume fraction equal to 13/18, see Eq. (2.16) in the text.

In turn, the simple reasoning allows finding a reliable evaluation. By analogy to the case with $k = 3$, we assume that for any odd scale $k$ only the above-mentioned HL and HH configurations have non-zero probabilities of appearance in the limit $\beta\Delta \to -\infty$. One can presume that a possible difference between odd and even $k$ is negligible for $k \gg 1$. For a given odd $k$-scale, to the HL-pattern the parameters mentioned earlier are as follows: $m_{HL} = 0$ and $n_{HL} = (k^2 + 1)/2$. In turn, for the HH-configuration the relative parameters are: $m_1 = 2k(k-1)$ and $n_{HH} = k^2$. Thus, one can get straightforwardly the formula for the limiting probability curve

$$p^*_H(k) = \begin{cases} 0, & 0 \leq \varphi < \frac{1}{2}\left(1 + \frac{1}{k^2}\right) \\ \dfrac{k^2(2\varphi - 1) - 1}{k^2 - 3}, & \frac{1}{2}\left(1 + \frac{1}{k^2}\right) \leq \varphi \leq 1 - \frac{1}{k^2} \\ 1, & 1 - \frac{1}{k^2} < \varphi \leq 1 \end{cases} \quad (2.17)$$

Consequently, using the condition given by Eq. (2.10), the limiting critical concentration is

$$\phi_c^*(k) = \frac{3}{4}\left(1 - \frac{1}{3k^2}\right). \tag{2.18}$$

Note that after substitution of $k = 3$ (the odd scale) to Eqs. (2.17-18) the same $\phi_c^*$-value as the resulting one from Eq. (2.16) is now recovered, that is 13/18.

For the purpose of illustration of the percolation behaviour at a fixed scale for Z4 and Z4s-models, in Fig. 3a we make a comparison between the evolving effective conductivity $\sigma^*(k = 3; Z4)$ described by Eq. (2.5), the solid lines, and the $\sigma^*(k = 3; Z4s)$ from Eq. (2.9), the dashed lines, for chosen interaction energies $\beta\Delta = 1, 0, -1$ and $-4$. As expected, the corresponding positions of the percolation threshold overlap despite the differences between the $\sigma^*(k = 3; Z4)$-values and the $\sigma^*(k = 3; Z4s)$ ones in the neighbourhood of critical concentrations. In addition, the black short-dashed line shows the limiting behaviour of $\sigma^*(k = 3; Z4s)$ for $\beta\Delta \to -\infty$. We underline again that the related critical concentration $\phi_c^*(k = 3) = 13/18$ marked by the vertical arrow in Fig. 3a comes from two different formulas: the exact one given by Eq. (2.16) and the reliable evaluation included in Eq. (2.18).

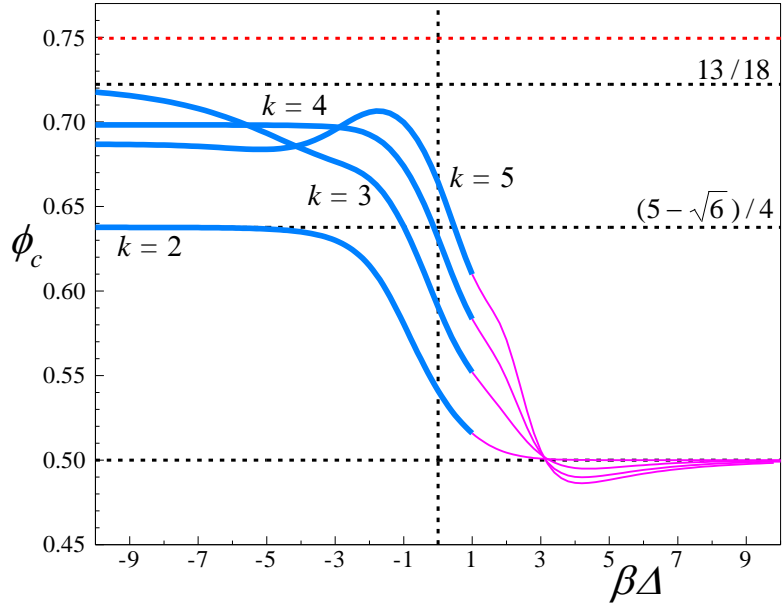

**Fig. 3b.** A family of critical concentration $(\phi(k; Z4s))_c$ curves for chosen length scales $k = 2, 3, 4$ and 5 as a function of interaction energy parameter $\beta\Delta$. These curves demonstrate peculiar features of the multi-scale percolation. The thick blue lines relate to the range of interest, $-\infty < \beta\Delta \leq 1$, where the system is a monophasic electric insulator/conductor depending on the volume fraction $\phi$, compare to Fig. 4 in [1]. The continuations of these curves nearly intersect at a characteristic point close to $\beta\Delta = 3$ (the thin rose lines) that belongs to the region where the system undergoes a phase separation. The case $\beta\Delta = 3$ could make a theme of future analyses. The middle two horizontal dashed lines correspond to limiting percolation thresholds when $\beta\Delta \to -\infty$ and they are determined from Eqs. (2.15) and (2.16), while the uppermost dashed red line relates to the estimated limit $\phi_c^*(k \gg 1) = 0.75$.

Now, we would like to present Fig. 3b which summarizes the main results. Here, at fixed length scales $k = 2, 3, 4$ and 5 the critical concentration $\phi_c(k; Z4s)$ as a function of interaction energy $\beta\Delta$ is depicted. These curves manifest unusual features of the multi-scale percolation for considered conditions. The thick (blue online) lines relate to the physically sound range, $-\infty < \beta\Delta \leq 1$. For $\phi < \phi_c(k; Z4s)$ we have low-conductivity $L$-phase and correspondingly, when $\phi > \phi_c(k; Z4s)$ the system becomes a $H$-conductor. For example, within the range which is taken into account, the thick curve for $k = 2$ relates to the percolation line marked by the open diamonds in Fig. 4 of Ref. [1]. However, the continuations of those thick curves for $\beta\Delta > 1$, represented by the thin rose lines, shrink to the characteristic point close to $\beta\Delta = 3$ and then, finally, converge from the bottom to the $\phi_c(k = 1; Z4s) = 0.5$. Nevertheless, the characteristic point belongs to the region, where the system undergoes a phase separation; *cf.* Fig. 4 of Ref. [1]. This point could offer a subject for further study in another context. The middle two horizontal dashed lines relate to limiting critical concentrations $\phi^*_c(k)$ at length scales $k = 2$ and 3, when $\beta\Delta \to -\infty$. They can be determined from the exact formulas given by Eqs. (2.15-16). The uppermost horizontal (red) line corresponds to the limiting critical concentration for larger length scales. When $k \gg 1$, the odd-even difference becomes negligible and from Eq. (2.18) we obtain the highest possible threshold value, i.e. $\phi^*_c(k) \cong 0.75$.

## 3. Conclusions

In this paper, we investigate the *multi-scale* percolation behaviour for the specific lattice model [1] developed for the case of interacting particle at scale $k = 2$, yet modified to any scale $k$, without interactions [2]. We consider the physically sound range of interaction energy (see the text), when either the repulsion or the attraction appear between adjacent conducting particles. When the interaction energy is weaker than the thermal energy, i.e. $|\beta\Delta| < 1$, then the behaviour of the multi-scale percolation is determined by thermodynamic fluctuations and geometrical features of local configurations $k \times k$. On the other hand, for $\beta\Delta \ll -1$ thermodynamic fluctuations are suppressed by the repulsive interaction energy. Then, this kind of interactions decides about the percolation behaviour.

Here, we use the so-called Z4-model accounting for all local conductivities and cluster energies as well as its simplified version the Z4s-model. The latter one reduces the whole number of local configurations to two classes. It was numerically established that the both approaches keep the same thresholds. However, using the Z4s-model, analytical formulas

13
at scales $k = 2$ and 3 can be obtained when the repulsive interaction energy dominates the thermal one.

Contrary to the already observed displacements of the percolation threshold within the range $0.5 \leq \phi_c(k) < 1$ caused mainly by the geometrical features [2], here the dominating repulsive interaction energy reduces the range to the following one: $0.5 \leq \phi_c(k) < \phi^*_c(k)$. The simple analogy to the case $k = 3$ allows evaluating the limiting probability of the appearance of *H*-configuration. In such a way, the analytical formulas for $p^*_H(k)$ and for the limiting critical concentration $\phi^*_c(k)$ can be obtained for any odd *k*. However, for any $k \gg 1$ the related parity differences become negligible. Thus, the highest possible value of the threshold is to be 0.75.